# PyUnfold: A Python package for iterative unfolding

**James Bourbeau**[1] **and Zigfried Hampel-Arias**[1, 2]

**1** University of Wisconsin - Madison, Wisconsin, USA **2** IIHE, Université Libre de Bruxelles, Bruxelles, Belgium

## Summary

In an ideal world, experimentalists would have access to the perfect detector: an apparatus that makes no error in measuring a desired quantity. However, in real life scenarios, detectors have:

- Finite resolutions
- Characteristic biases that cannot be eliminated
- Less than 100% detection efficiencies
- Statistical uncertainties

By building a matrix that encodes a detector's smearing of the desired true quantity into the measured observable(s), a deconvolution can be performed that provides an estimate of the true variable. This deconvolution process is known as unfolding.

PyUnfold is an extensible framework for the unfolding of discrete probability distributions via the iterative unfolding method described in (D'Agostini 1995). This method falls into the general class of inverse problems, and is especially powerful when no analytic form of the inverse function is explicitly available, instead requiring an estimate (e.g. via a finite amount of simulation) of the response function. Given that measured data comprise a finite sample, PyUnfold also implements the uncertainty contributions stated in (Adye 2011).

The unfolding method itself is data-agnostic, referring to the measurement process as the smearing of a set of true causes into a set of detectable effects. For example one could define as causes the true energy of a particle and the effects the measured energy of that particle in a detector. Another example might be a set of diseases (causes) and possible clinical symptoms (effects). So long as it is possible to encode estimable resolutions and biases connecting causes to effects in a binned response matrix, one can perform a deconvolution with PyUnfold.

The primary purpose of PyUnfold is to provide an unfolding toolkit for members of all scientific disciplines in an easy-to-use package. For example, unfolding methods are commonly used in the analysis pipeline of the high-energy physics (HEP) community. However, the deconvolution packages used in HEP maintain a strong dependence on the ROOT data analysis framework (ROOT 1997), which is almost exclusively used in HEP. Instead, PyUnfold is built on top of the Python scientific computing stack (i.e. NumPy, SciPy, and pandas), thus broadening its scope to a general scientific audience.

PyUnfold has been designed to be both easy to use for first-time users as well as flexible enough for fine-tuning an analysis and seamlessly testing the robustness of results. It provides support for the following:

- Custom, user defined initial prior probability distributions, the default being the uniform prior. The non-informative Jeffreys' prior (Jeffreys 1946) is accessible as a utility function.

- Unfolding stopping criteria based on test statistic calculations comparing unfolded distributions from one iteration to the next. These include Kolmogorov-Smirnov (Kolmogorov 1933)(Smirnov 1948), $\chi^2$, relative difference, and Bayes factor (BenZvi et al. 2011) tests.
- Tunable spline regularization as a means of ensuring that unfolded distributions do not suffer from growing fluctuations potentially arising from the finite binning of the response matrix.
- Option to choose between Poisson or multinomial forms of the covariance matrices for both the data and response contributions to the uncertainty calculation.
- Multivariate unfolding via definitions of subsets of causes, which are regularized in their respective blocks or groups.

Further mathematical details regarding the iterative unfolding procedure, including complete derivations of the statistical and systematic uncertainty propagation can be found in the online documentation.

PyUnfold has been applied successfully for the measurement of the cosmic-ray energy spectrum by the HAWC Observatory (Alfaro 2017) and is currently being used in an analysis by members of the IceCube Neutrino Observatory.

# Acknowledgements

The authors acknowledge support from the Wisconsin IceCube Particle Astrophysics Center at the UW-Madison, and especially for the guidance of Professor Stefan Westerhoff. We also acknowledge the financial support provided by the National Science Foundation, the Belgian American Educational Foundational Fellowship, and Wallonie-Bruxelles International.

Smirnov, N. 1948. "Table for Estimating the Goodness of Fit of Empirical Distributions." *Ann. Math. Statist.* 19 (2). The Institute of Mathematical Statistics:279–81. https://doi.org/10.1214/aoms/1177730256.

# ITERATIVE UNFOLDING REFERENCE

Towards understanding iterative unfolding

May 16, 2018

Revision 1.0

# Iterative Unfolding Reference

James Bourbeau, Zigfried Hampel-Arias

May 16, 2018

## Abstract

This reference outlines the D'Agostini procedure used for iterative unfolding. We first motivate and outline the method in a manner suitable for users new to the technique. Next is shown the full propagation of errors due to the unfolding process, including the derivation of the final form of the covariance matrix.

# Contents

# 1 Introduction

The general class of unfolding methods is amongst the physicist's toolbox as a powerful means to connect an experiment's observable variables with true physical quantities. Typically a matrix can be built to encompass the effects of the measurement process on a simulated 'true' distribution and the manifestation of said distribution as an experimenter's desired observable. With this response matrix, a distribution of the observable in an experiment can be **unfolded**, providing an estimate of the true parent distribution.

A variety of unfolding methods exist, each with its respective strengths and weaknesses. For example, the simplest method is the matrix inversion unfolding, which for a well populated, highly linear response matrix can be both efficient and precise. However, even with relatively small off-diagonal elements, this method can be unfavorable, as the matrix may be singular or may introduce wildly fluctuating results due to limited statistics. There exist methods to quell such issues, though these require the tuning of various parameters which typically have no physical connection to the experiment at hand.

Here we discuss D'Agostini's iterative unfolding technique presented in [1], a manifestly inferential method. Starting from Bayes' theorem, an iterative unfolding procedure is developed, which then can be implemented without too much difficulty for the typical experimenter. This document has been adapted from Chapter 7 and Appendix B of [2].

## 2 Method

As discussed in the Section 1, the conceptually simplest way to connect true causes $C_\mu$ and observable effects $E_j$ is via their respective count distributions, $n$ and $\phi$, a reponse matrix, $R$, and it's inverse $M$[1]:

$$n(E) = R\,\phi(C),$$
$$\phi(C) = M\,n(E). \qquad (1)$$

Due to the aforementioned potential difficulties in matrix inversion, we can take into consideration Bayes' theorem,

$$P(C_\mu|E_j) = \frac{P(E_j|C_\mu)\,P(C_\mu)}{\sum_\nu^{n_C} P(E_j|C_\nu)\,P(C_\nu)}, \qquad (2)$$

where $n_C$ is the number of possible causes. Equation 2 dictates that having observed the effect $E_j$, the probability that it's origin is due to the cause $C_\mu$ is proportional to product of the probability of the cause and the probability of the cause to produce that effect. Hence, the elements $P(E_i|C_\mu)$ represent the probability that a given $C_\mu$ results in the effect $E_i$, and is the response matrix typically generated via modeling or simulation. Continuing with $P(C_\mu|E_j)$, we can then connect the measured observed effects to their causes by

$$\phi(C_\mu) = \sum_i^{n_E} P(C_\mu|E_i)\,n(E_i). \qquad (3)$$

Stepping back to eq. 2 for a moment, one identifies $P(C_\mu)$ as the prior cause distribution, representing our current knowledge of the causes. The prior is a normalized distribution such that $\sum_\mu^{n_C} P(C_\mu) = 1$. This normalization requirement is not imposed on the response matrix efficiency $\epsilon_\mu$: $0 \leq \epsilon_\mu = \sum_j^{n_E} P(E_j|C_\mu) \leq 1$, ie, a cause does not need to produce any effect. Taking this (in)-efficiency into account, we rewrite eq. 3 as

$$\phi(C_\mu) = \frac{1}{\epsilon_\mu} \sum_i^{n_E} P(C_\mu|E_i)\,n(E_i). \qquad (4)$$

Identifying here the explicit form of $M$, the full matrix (Bayesian) inversion equation is then

$$\phi(C_\mu) = \sum_j^{n_E} M_{\mu j}\,n(E_j), \qquad (5)$$

where

$$M_{\mu j} = \frac{P(E_i|C_\mu)\,P(C_\mu)}{[\sum_k^{n_E} P(E_k|C_\mu)][\sum_\nu^{n_C} P(E_i|C_\nu)\,P(C_\nu)]}. \qquad (6)$$

The response matrix $P(E_i|C_\mu)$ is generated via simulation, and the $n(E_i)$ provided through measurement, apparently bestowing the freedom to choose the form of $P(C_\mu)$. Again, $P(C_\mu)$ represents the total of our prior knowledge of the parent distribution. Typically an experimenter refrains from introducing bias in the prior, and two such appropriate choices are the uniform and non-informative Jeffreys' [3] priors:

$$P_{\text{Uniform}}(C_\mu) = \frac{1}{n_C}$$
$$P_{\text{Jeffrey}}(C_\mu) = \frac{1}{\log(C_{max}/C_{min})\, C_\mu},$$

keeping in mind that the these priors place the causes on equal footing, not that all parent cause distributions are equally probable.

---

[1] Except for C and E, all variables and subscripts related to causes are Greek letters, while Latin letters are used for effects. The only superscript is the iteration number, i.

We now possess all the necessary machinery to perform an unfolding. Having started with a conservative initial prior, the unfolded result is a Bayesian best estimate of the true distribution. There is nothing stopping us from using this result as the best knowledge estimate of $P(C_\mu)$ in eq. 6 for a subsequent unfolding. We can take this any number of steps further, making the process an iterative unfolding. Thus, after calculating $\phi(C_\mu)$ via eq. 5, we recalculate $M_{\mu j}$ per eq. 6, returning again to eq. 5 for an updated $\phi(C_\mu)'$. Since $P(C_\mu) = \frac{\phi_\mu}{\sum_\nu \phi_\nu} = \frac{\phi_\mu}{N_{\text{True}}}$, where $N_{\text{True}}$ is the estimated true number of cause events, we can make the change $P(C_\mu) \to \phi_\mu$ in eq. 6. Adding the iteration superscript and shortening the notation[1], this equates to

$$M_{\mu j} = \frac{P_{\mu j}\, \phi_\mu^i}{\epsilon_\mu \sum_\rho P_{\rho j}\, \phi_\rho^i}$$
$$\phi_\mu^{i+1} = \sum_j M_{\mu j}\, n_j.$$

The unfolding proceeds until a desired stopping criterion is satisfied, say by comparing subsequent iterations with a test statistic such as a $\chi^2$. The algorithm below outlines the basics to the iterative unfolding scheme:

---
**Algorithm 1** Unfolding Algorithm
---
$\phi^0 \leftarrow$ Prior
testStatistic$\leftarrow$ Pass
**while** ( testStatistic = Pass ) **do**
    $M \leftarrow M(P(E|C), \phi^i)$
    $\phi^{i+1} \leftarrow M \times n$
    testStatistic$\leftarrow$ TS$(\phi^i, \phi^{i+1})$
**end while**

## 3 Regularization

After each iteration, the resulting posterior distribution, $P(C_\mu)$, is our new best guess of the (normalized) parent distribution. Using this best estimate as the prior for the next iteration, one can induce large fluctuations in neighboring $C_\mu$ bins. It is here the equivalence of matrix inversion techniques and iterative unfolding is seen. After many iterations, wild fluctuations can appear, indicating the granularity in the MC derived $P_{\mu j}$. Furthermore, in using the posterior as the subsequent prior, one is 'telling' the unfolding that physical distributions of that nature are allowable priors. Instead, as pointed out in [1] (section 6.3), for an experimenter interested in a particular model's parameters, fitting all but the last posterior is equivalent to performing a maximum likelihood fit to the data.

As physical measurements are expected to be smooth (a safe assumption for energy spectra for example), one can regularize the $\phi_\mu^i$. In principle one can choose any smoothing function. For the cosmic-ray energy spectrum for example, $\phi_\mu^i$ can be simply fit to a power law or a spline as was done in [4], using the fitted function as the input prior for the next iteration. While this could be seen as a loss of information, it is important to remember that **any** improved prior distribution will enhance our estimation method, along with the **prior** expectation that our distribution is smooth.

The other possibility is to avoid regularization altogether and instead ensure that $P_{\mu j}$ is smooth enough. The granularity of the cause and effect bins will dictate the degree of smoothness required to ensure non-fluctuating $\phi^i$ solutions. The more widely used techniques for smoothing $P_{\mu j}$ include kernel density estimation and penalized spline fitting routines.

# 4 Unfolding Uncertainties

To begin the excursion into the calculation of uncertainties, we first shorten the notation in accordance with footnote [1]:

$$P(E_i|C_\mu) = P_{\mu i} \qquad \phi(C_\mu) = \phi_\mu \qquad n(E_j) = n_j.$$

As outlined in [1] (section 4), the covariance matrix $V = V(\phi, \phi')$ from statistical contributions has two components: $V^{\text{Data}}$ from the counted measured effects distribution, and $V^{\text{MC}}$ due to the limited MC statistics in $P_{\mu j}$. This can be seen from considering the uncertainties from $n_j$ and $M_{\mu j}$ in eq. 5. Since $\phi = M \times n = M(P(E|C)) \times n$, we can identify respectively the aforementioned error contributions as

$$\begin{aligned} V^{Total} &= V^{Data} + V^{MC} \\ &= \frac{\partial \phi}{\partial n} Cov(n,n') \frac{\partial \phi'}{\partial n} \\ &+ \frac{\partial \phi}{\partial P} Cov(P,P') \frac{\partial \phi'}{\partial P}. \end{aligned}$$

## 4.1 $V^{Data}$

D'Agostini argues that since the data sample $n_j$ is a realization of a multinomial distribution, then

$$V^{Data} = M\, Cov(n,n')\, M \tag{7}$$

where the $Cov(n,n')$ is the covariance matrix of the measurements with respect to the estimated true number of events $\sum_\mu \phi_\mu = N_{true}$:

$$Cov(n_k, n_j) = \begin{cases} n_j(1 - \frac{n_j}{N_{true}}) & \text{if } k = j \\ -\frac{n_j n_k}{N_{true}} & \text{if } k \neq j \end{cases}. \tag{8}$$

However, Adye ([5] section 5) demonstrates that this error estimation is only valid for the first iteration, as subsequent $\phi^i$ are **not independent** of $n_j$. Indeed, we should re-write eq. 7 appropriately as

$$V^{Data} = \frac{\partial \phi^{i+1}}{\partial n} \times Cov(n,n') \times \frac{\partial \phi^{i+1\prime}}{\partial n}, \tag{9}$$

with

$$\frac{\partial \phi_\mu^{i+1}}{\partial n_j} = M_{\mu j} + \frac{\phi_\mu^{i+1}}{\phi_\mu^i} \frac{\partial \phi_\mu^i}{\partial n_j} - \sum_{\sigma,k} \epsilon_\sigma \frac{n_k}{\phi_\sigma^i} M_{\mu k} M_{\sigma k} \frac{\partial \phi_\sigma^i}{\partial n_j}$$

where again the superscripts $i$ and $i+1$ refer to the iteration number. The full derivation of $\frac{\partial \phi^{i+1}}{\partial n}$ (eq. 20) is found in section 4.4.2 below. In some cases it is safe to use the Poisson form of $Cov(n,n')$:

$$Cov(n_k, n_j) = n_k\, \delta_{kj}. \tag{10}$$

## 4.2 $V^{MC}$

The contribution from $V^{MC}$, while well outlined in [1] and below, is quite a monster. If one simply implements the equation verbatim into code, the expected time for calculating all elements $\sim$ (number of bins)$^7$. Thus, here we present the form of $V^{MC}$, while in section 4.4.3 we show the explicit expansion and further contraction of indices towards a more reasonable, practical calculation.

D'Agostini identifies $V^{MC}$ via $\frac{\partial}{\partial M}$ giving

$$V^{MC} = n \times Cov(M, M') \times n'. \tag{11}$$

Further expansion reveals

$$Cov(M_{\mu k}, M_{\lambda j}) = \sum_{\{\sigma r\}, \{\sigma s\}} \frac{\partial M_{\mu k}}{\partial P_{\sigma r}} \frac{\partial M_{\lambda j}}{\partial P_{\sigma s}} Cov(P_{\sigma r}, P_{\sigma s}), \tag{12}$$

$$\frac{\partial M_{\mu k}}{\partial P_{\sigma j}} = M_{\mu k} \left[ \frac{\delta_{\mu\sigma} \delta_{jk}}{P_{\sigma j}} - \frac{\delta_{\mu\sigma}}{\epsilon_\sigma} - \frac{\delta_{jk} M_{\sigma k} \epsilon_\sigma}{P_{\sigma k}} \right], \tag{13}$$

$$Cov(P_{\sigma r}, P_{\sigma s}) = \begin{cases} \frac{1}{\tilde{n}_\sigma} P_{\sigma r} (1 - P_{\sigma r}) & \text{if } r = s \\ -\frac{1}{\tilde{n}_\sigma} P_{\sigma r} P_{\sigma s} & \text{if } r \neq s \end{cases}. \tag{14}$$

In the final expression, $\tilde{n}_\mu$ represents the number of simulated events which fell into the true cause bin $\mu$. If our simulation is weighted, we identify $\tilde{n}$ with the effective number of events $\tilde{n}_\mu = \frac{(\sum_j w_{\mu j})^2}{\sum_j w_{\mu j}^2}$ for all $j$ events in bin $\mu$.

Once again, Adye ([6]) shows this is a first order estimate, only valid for the first iteration. Re-writing 11 with $\frac{\partial}{\partial P}$,

$$V^{MC} = \frac{\partial \phi^{i+1}}{\partial P} \times Cov(P, P') \times \frac{\partial \phi^{i+1'}}{\partial P}, \tag{15}$$

we identify $\frac{\partial \phi^{i+1}}{\partial P}$ as

$$\frac{\partial \phi_\mu^{i+1}}{\partial P_{\lambda k}} = \frac{\delta_{\lambda \mu}}{\epsilon_\mu} \left( \frac{n_k \phi_\mu^i}{f_k} - \phi_\mu^{i+1} \right) - \frac{n_k \phi_\lambda^i}{f_k} M_{\mu k}$$
$$+ \frac{\phi_\mu^{i+1}}{\phi_\mu^i} \frac{\partial \phi_\mu^i}{\partial P_{\lambda k}} - \sum_{\rho, j} n_j \frac{\epsilon_\rho}{\phi_\rho^i} M_{\rho j} M_{\mu j} \frac{\partial \phi_\rho^i}{\partial P_{\lambda k}}$$

whose derivation (eq. 21) is found in section 4.4.3 below. Of course, D'Agostini's form of $Cov(P, P')$ remains valid for use with the new construction of the partials. One may also use a Poisson covariance if justified appropriately:

$$Cov(P_{\rho r}, P_{\lambda s}) = \sigma_{\rho r} \sigma_{\lambda s} \delta_{\rho \lambda} \delta_{rs}, \tag{16}$$

with $\sigma_{\rho r}$ being the error estimates on $P_{\rho r}$ estimated when filling $P$ with Monte Carlo.

**Algorithm 2** Unfolding Algorithm - Including Errors

$\phi^0 \leftarrow$ Prior
testStatistic$\leftarrow$ Pass
**while** ( testStatistic = Pass ) **do**
    $M \leftarrow M(P(E|C), \phi^i)$
    $\phi^{i+1} \leftarrow M \times n$
    $\frac{\partial \phi^{i+1}}{\partial n} \leftarrow$ eq. 20
    $\frac{\partial \phi^{i+1}}{\partial P} \leftarrow$ eq. 21
    testStatistic$\leftarrow$ TS$(\phi^i, \phi^{i+1})$
**end while**
$V^{Total} \leftarrow V^{Data}(\frac{\partial \phi^{i+1}}{\partial n}) + V^{MC}(\frac{\partial \phi^{i+1}}{\partial P})$
$\sigma_\phi^2 \approx diag(V^{Total})$

### 4.3 Updated Unfolding Algorithm

The afore-outlined unfolding algorithm must be modified to include the propagation of systematic errors. At each iteration we have $\phi^{i+1}$, so both $\frac{\partial \phi^{i+1}}{\partial n}$ and $\frac{\partial \phi^{i+1}}{\partial P}$ can be calculated. The results are propagated and saved until the full covariance matrix is required for error estimates on the final $\phi$.

### 4.4 Expansion of Components of $V$

#### 4.4.1 Some useful formulae

Recalling the unfolding formulae from before,

$$\phi_\mu^{i+1} = \sum_k M_{\mu k}\, n_k \qquad\qquad M_{\mu j} = \frac{P_{\mu j}\, \phi_\mu^i}{\epsilon_\mu\, f_j},$$

where the efficiency, $\epsilon$, and normalization, $f$, of $M$ are

$$\epsilon_\mu = \sum_j P_{\mu j} \qquad\qquad f_j = \sum_\mu P_{\mu j}\, \phi_\mu^i.$$

Of note is the presence of $\phi^i$, ie, the unfolded cause distribution from the previous iteration, or the prior in the case $i = 0$.

We will be taking derivatives of these objects with respect to $n_k$ and $P_{\lambda k}$, to wit,

$$\frac{\partial P_{\mu j}}{\partial n_k} = 0 \qquad \frac{\partial \epsilon_\mu}{\partial n_k} = 0 \qquad \frac{\partial f_j}{\partial n_k} = \sum_\mu P_{\mu j} \frac{\partial \phi_\mu^i}{\partial n_k} \qquad (17)$$

$$\frac{\partial P_{\mu j}}{\partial P_{\lambda k}} = \delta_{\mu\lambda}\, \delta_{jk} \qquad \frac{\partial \epsilon_\mu}{\partial P_{\lambda k}} = \delta_{\lambda\mu} \qquad \frac{\partial f_j}{\partial P_{\lambda k}} = \delta_{jk}\, \phi_\lambda^i + \sum_\mu P_{\mu j} \frac{\partial \phi_\mu^i}{\partial P_{\lambda k}}. \qquad (18)$$

The explicit forms of $\frac{\partial \phi_\mu^i}{\partial n_k}$ and $\frac{\partial \phi_\mu^i}{\partial P_{\lambda k}}$ will be shown below, but only for $i = 0$ do

$$\frac{\partial \phi_\mu^i}{\partial n_k} = 0 \quad, \quad \frac{\partial \phi_\mu^i}{\partial P_{\lambda k}} = 0, \qquad (19)$$

as no unfolding has been performed. This will clearly not be the case for subsequent iterations when $\phi^i$ becomes dependent on $n_k$ and $P_{\lambda k}$.

### 4.4.2 Expansion of $V^{Data}$

Making the appropriate substitutions, the index representation of eq. 9 is

$$V(\phi_\mu^{i+1}, \phi_\nu^{i+1})^{Data} = \sum_{jk} \frac{\partial \phi_\mu^{i+1}}{\partial n_j} Cov(n_j, n_k) \frac{\partial \phi_\nu^{i+1}}{\partial n_k},$$

with

$$\begin{aligned}
\frac{\partial \phi_\mu^{i+1}}{\partial n_j} &= \frac{\partial}{\partial n_j} \sum_k M_{\mu k} n_k \\
&= \sum_k (M_{\mu k} \frac{\partial n_k}{\partial n_j} + n_k \frac{\partial M_{\mu k}}{\partial n_j}) \\
&= \sum_k (M_{\mu k} \delta_{jk} + n_k \frac{\partial M_{\mu k}}{\partial n_j}) \\
&= M_{\mu j} + \sum_k n_k \underbrace{\frac{\partial M_{\mu k}}{\partial n_j}}
\end{aligned}$$

$$\begin{aligned}
\frac{\partial M_{\mu k}}{\partial n_j} &= \frac{\partial}{\partial n_j} \frac{P_{\mu k} \phi_\mu^i}{\epsilon_\mu f_k} \\
&= \underbrace{\frac{P_{\mu k}}{\epsilon_\mu f_k}}_{\frac{M_{\mu k}}{\phi_\mu^i}} \frac{\partial \phi_\mu^i}{\partial n_j} - \underbrace{\frac{P_{\mu k} \phi_\mu^i}{\epsilon_\mu f_k}}_{M_{\mu k}} \frac{1}{f_k} \sum_\sigma P_{\sigma k} \frac{\partial \phi_\sigma^i}{\partial n_j} \\
&= \frac{M_{\mu k}}{\phi_\mu^i} \frac{\partial \phi_\mu^i}{\partial n_j} - M_{\mu k} \sum_\sigma \epsilon_\sigma \underbrace{\frac{P_{\sigma k}}{\epsilon_\sigma f_k}}_{\frac{M_{\sigma k}}{\phi_\sigma^i}} \frac{\partial \phi_\sigma^i}{\partial n_j} \\
&= \frac{M_{\mu k}}{\phi_\mu^i} \frac{\partial \phi_\mu^i}{\partial n_j} - \sum_\sigma \frac{\epsilon_\sigma}{\phi_\rho^i} M_{\mu k} M_{\sigma k} \frac{\partial \phi_\sigma^i}{\partial n_j}
\end{aligned}$$

$$\frac{\partial \phi_\mu^{i+1}}{\partial n_j} = M_{\mu j} + \frac{1}{\phi_\mu^i} \frac{\partial \phi_\mu^i}{\partial n_j} \underbrace{\sum_k M_{\mu k} n_k}_{\phi_\mu^{i+1}} - \sum_{\sigma,k} \epsilon_\sigma \frac{n_k}{\phi_\rho^i} M_{\mu k} M_{\sigma k} \frac{\partial \phi_\sigma^i}{\partial n_j}$$

$$\frac{\partial \phi_\mu^{i+1}}{\partial n_j} = M_{\mu j} + \frac{\phi_\mu^{i+1}}{\phi_\mu^i} \frac{\partial \phi_\mu^i}{\partial n_j} - \sum_{\sigma,k} \epsilon_\sigma \frac{n_k}{\phi_\sigma^i} M_{\mu k} M_{\sigma k} \frac{\partial \phi_\sigma^i}{\partial n_j} \qquad (20)$$

Recalling eq. 19, $\frac{\partial \phi_\mu^0}{\partial n_j} = 0$ for the first iteration, eliminating the last two terms of eq. 20 and recovering $\frac{\partial \phi_\mu^1}{\partial n_j} = M_{\mu j}$ as per [1]. In practice, one need only calculate $\frac{\partial \phi_\mu^{i+1}}{\partial n_j}$ for each iteration, saving the result until the full calculation of $V^{Data}$ is required.

### 4.4.3 Expansion of $V^{MC}$

Similar to $V(\phi_\mu^{i+1}, \phi_\nu^{i+1})^{Data}$, we identify the contributions to $V$ from the Monte Carlo:

$$V(\phi_\mu^{i+1}, \phi_\nu^{i+1})^{MC} = \sum_{\lambda j} \sum_{\rho k} \frac{\partial \phi_\mu^{i+1}}{\partial P_{\lambda j}} Cov(P_{\lambda j}, P_{\rho k}) \frac{\partial \phi_\nu^{i+1}}{\partial P_{\rho k}}.$$

Proceeding forward,

$$\frac{\partial \phi_\mu^{i+1}}{\partial P_{\lambda k}} = \frac{\partial}{\partial P_{\lambda k}} \sum_j M_{\mu j} n_j = \sum_j n_j \underbrace{\frac{\partial M_{\mu j}}{\partial P_{\lambda k}}}$$

$$\begin{aligned}
\frac{\partial M_{\mu j}}{\partial P_{\lambda k}} &= \frac{\partial}{\partial P_{\lambda k}} \frac{P_{\mu j} \phi_\mu^i}{\epsilon_\mu f_j} \\
&= \frac{\phi_\mu^i}{\epsilon_\mu f_j} \frac{\partial P_{\mu j}}{\partial P_{\lambda k}} + \underbrace{\frac{P_{\mu j}}{\epsilon_\mu f_j}}_{\frac{M_{\mu j}}{\phi_\mu^i}} \frac{\partial \phi_\mu^i}{\partial P_{\lambda k}} - \frac{1}{\epsilon_\mu f_j} \underbrace{\frac{P_{\mu j} \phi_\mu^i}{\epsilon_\mu f_j}}_{M_{\mu j}} \left( f_j \frac{\partial \epsilon_\mu}{\partial P_{\lambda k}} + \epsilon_\mu \frac{\partial f_j}{\partial P_{\lambda k}} \right) \\
&= \frac{\phi_\mu^i}{\epsilon_\mu f_j} \delta_{\lambda \mu} \delta_{jk} + \frac{M_{\mu j}}{\phi_\mu^i} \frac{\partial \phi_\mu^i}{\partial P_{\lambda k}} - \frac{1}{\epsilon_\mu f_j} M_{\mu j} \left( f_j \delta_{\lambda \mu} + \epsilon_\mu \delta_{jk} \phi_\lambda^i + \epsilon_\mu \sum_\rho P_{\rho j} \frac{\partial \phi_\rho^i}{\partial P_{\lambda k}} \right) \\
&= \frac{\phi_\mu^i}{\epsilon_\mu f_j} \delta_{\lambda \mu} \delta_{jk} + \frac{M_{\mu j}}{\phi_\mu^i} \frac{\partial \phi_\mu^i}{\partial P_{\lambda k}} - \frac{M_{\mu j}}{\epsilon_\mu} \delta_{\lambda \mu} - \frac{M_{\mu j} \phi_\lambda^i}{f_j} \delta_{jk} - \sum_\rho \epsilon_\rho M_{\mu j} \underbrace{\frac{P_{\rho j}}{\epsilon_\rho f_j}}_{\frac{M_{\rho j}}{\phi_\rho^i}} \frac{\partial \phi_\rho^i}{\partial P_{\lambda k}} \\
&= \frac{\phi_\mu^i}{\epsilon_\mu f_j} \delta_{\lambda \mu} \delta_{jk} + \frac{M_{\mu j}}{\phi_\mu^i} \frac{\partial \phi_\mu^i}{\partial P_{\lambda k}} - \frac{M_{\mu j}}{\epsilon_\mu} \delta_{\lambda \mu} - \frac{M_{\mu j} \phi_\lambda^i}{f_j} \delta_{jk} - \sum_\rho M_{\rho j} M_{\mu j} \frac{\epsilon_\rho}{\phi_\rho^i} \frac{\partial \phi_\rho^i}{\partial P_{\lambda k}},
\end{aligned}$$

and going back to $\frac{\partial \phi_\mu^{i+1}}{\partial P_{\lambda k}}$ to include the sum over $j$,

$$\frac{\partial \phi_\mu^{i+1}}{\partial P_{\lambda k}} =$$

$$\sum_j n_j \Big[ \frac{\phi_\mu^i}{\epsilon_\mu f_j}\delta_{\lambda\mu}\delta_{jk} + \frac{M_{\mu j}}{\phi_\mu^i}\frac{\partial \phi_\mu^i}{\partial P_{\lambda k}} - \frac{M_{\mu j}}{\epsilon_\mu}\delta_{\lambda\mu} - \frac{M_{\mu j}\phi_\lambda^i}{f_j}\delta_{jk} - \sum_\rho M_{\rho j} M_{\mu j} \frac{\epsilon_\rho}{\phi_\rho^i}\frac{\partial \phi_\rho^i}{\partial P_{\lambda k}} \Big]$$

$$= \frac{n_k \phi_\mu^i}{\epsilon_\mu f_k}\delta_{\lambda\mu} + \frac{1}{\phi_\mu^i}\frac{\partial \phi_\mu^i}{\partial P_{\lambda k}} \underbrace{\sum_j M_{\mu j} n_j}_{\phi_\mu^{i+1}} - \frac{\delta_{\lambda\mu}}{\epsilon_\mu}\underbrace{\sum_j M_{\mu j} n_j}_{\phi_\mu^{i+1}} - \frac{n_k M_{\mu k}\phi_\lambda^i}{f_k}$$

$$- \sum_j \sum_\rho n_j \frac{\epsilon_\rho}{\phi_\rho^i} M_{\rho j} M_{\mu j} \frac{\partial \phi_\rho^i}{\partial P_{\lambda k}},$$

with final form

$$\begin{aligned}
\frac{\partial \phi_\mu^{i+1}}{\partial P_{\lambda k}} &= \frac{\delta_{\lambda\mu}}{\epsilon_\mu}\Big( \frac{n_k \phi_\mu^i}{f_k} - \phi_\mu^{i+1} \Big) - \frac{n_k \phi_\lambda^i}{f_k} M_{\mu k} \\
&+ \frac{\phi_\mu^{i+1}}{\phi_\mu^i}\frac{\partial \phi_\mu^i}{\partial P_{\lambda k}} - \sum_{\rho,j} n_j \frac{\epsilon_\rho}{\phi_\rho^i} M_{\rho j} M_{\mu j} \frac{\partial \phi_\rho^i}{\partial P_{\lambda k}}.
\end{aligned} \quad (21)$$

Again for the first iteration $\frac{\partial \phi_\mu^0}{\partial P_{\lambda k}} = 0$, eliminating the last two terms of eq. 21, and recovering D'Agostini's version. Again, upon implementation one need only calculate $\frac{\partial \phi_\mu^{i+1}}{\partial P_{\lambda k}}$ at each iteration, saving it until $V^{MC}$ is needed for error estimation.